\documentclass[fleqn,twoside]{article}
\usepackage{espcrc2}

\usepackage{graphicx}
\usepackage[figuresright]{rotating}

\usepackage{euscript}

\newcommand{\Peu}{\EuScript{P}}
\newcommand{\AmS}{{\protect\the\textfont2
  A\kern-.1667em\lower.5ex\hbox{M}\kern-.125emS}}

\title{
\vspace{-20mm}
\begin{flushright} \bf IFJPAN-IV-2010-4\\ \end{flushright}
\vspace{5mm}
Monte Carlo modelling of NLO DGLAP QCD evolution\\
       in the fully unintegrated form}

\author{%
S. Jadach%
\address[IFJ]{
  Institute of Nuclear Physics, Polish Academy of Sciences,\\
  ul. Radzikowskiego 152, 31-342, Krak\'ow, Poland}%
\address[CERN]{Theory Group, Physics Department, 
  CERN, CH-1211, Geneva 23, Switzerland }%
,
A.~Kusina\addressmark[IFJ],
M.~Skrzypek\addressmark[IFJ]
and
M.~Slawinska\addressmark[IFJ]%
}

\begin{document}

\begin{abstract}
We are reporting on the ongoing effort of the
Monte Carlo (MC) modelling of NLO DGLAP QCD evolution
in the fully unintegrated form.
The resulting parton shower MC is performing on its own 
the NLO QCD evolution, contrary to all known programs
of this kind which are limited to LO level only.
We overview this new MC scheme, for the non-singlet
subset of the gluonstrahlung diagrams.
Precision numerical test of this new scheme are also shown.
\vspace{1pc}
\end{abstract}


\maketitle

\section{Introduction}
In the so called factorization theorems,
see for instance~\cite{Ellis:1978ty,Collins:1984kg,Bodwin:1984hc},
the scattering process with a single large transverse momentum
scale (short distance interaction),
 can be described
in the perturbative QCD~\cite{Gross:1973ju,Gross:1974cs,Georgi:1951sr}
(pQCD) in terms of the on-shell hard process
matrix element convoluted with the {\em ladder part}.
The hard process is calculable up to a fixed perturbative order.
The ladder part, calculated for each coloured parton entering (exiting)
the hard process, is conveniently described as a tree-like
stream of partons, parametrized by the inclusive parton distribution
function, PDF.

The logarithmic response of the inclusive PDF to the change of the
large $k_T$ scale is the so called DGLAP\cite{DGLAP} evolution of the PDFs.
This evolution was mastered for the inclusive PDFs up to NLO level
in the early 80's, see for instance ~\cite{Floratos:1978ny,Curci:1980uw},
and recently even to the NNLO level~\cite{Vogt:2004mw}.

Alternatively the parton tree encapsulated in the inclusive PDF can be
modelled using direct stochastic simulation of their four-momenta
and other attributes, the so-called Monte Carlo (MC) technique.
Here, the breakthrough was made in mid-80's, 
see refs.~\cite{Sjostrand:1985xi,Webber:1984if},
where the LO ladder was implemented in the so-called parton shower (PS) MCs.

Standard LO level PSMC implements also 
the hadronization of quark and gluon partons
into hadrons and is a workhorse in the software in all collider experiments.
The advent of the LHC puts a challenging requirement on the quality
of the pQCD calculations needed for the experimental data analysis.
Some of them will soon require that the PSMC is upgraded to the complete
NLO level.
This, however is not an easy task, mainly because classic factorization
theorems~\cite{Ellis:1978ty,Collins:1984kg,Bodwin:1984hc}
are not well suited for the exclusive MC implementation, but rather for
inclusive PDFs.

In this contribution we would like to report on the serious and successful
attempt of solving the above problem of constructing NLO PSMC.
We shall overview main technical points and present some numerical
tests of the scheme.

Possible profits from NLO PSMC include:
(a)~Complete set of ``unitegrated soft counterterms'' for combining
hard process ME at NNLO with NLO PSMC;
(b)~Natural extensions towards BFKL/CCFM at low $x$;
(c)~Better modelling of low scale phenomena, $Q<10GeV$,
quark thresholds, primordial $k^T$, underlying event, etc.;
(d)~Porting information on the parton distributions from DIS (HERA)
to W/Z/DY (LHC) in the MC itself,
instead in the collinear PDFs (universality must be preserved) 
and more.

The MC modelling of NLO DGLAP is not so much the aim itself --
it will be rather a starting platform for many
interesting developments in pQCD in many directions.

Presently, we concentrate on constructing NLO PSMC for QCD
initial state radiation (ISR) just for one initial parton such that:
(a)~ it is based on the collinear factorization theorems
\cite{Ellis:1978ty,Collins:1984kg,Bodwin:1984hc}
as rigorously as possible,
(b)~we take scheme of Curci-Furmanski-Petronzio (CFP) as a main reference
and guide (axial gauge, $\overline{MS}$ dimensional regularization),
(c)~NLO DGLAP evolution is reproduced exactly at the inclusive level
(c)~MC performs NLO evolution by itself,
using new exclusive NLO evolution kernels,
without help of external pretabulated inclusive PDFs.

We refer the reader to 
Refs.~\cite{Jadach:2010ew,Jadach:2009gm,Kusina:2010gp,Slawinska:2009gn}
for discussions of other aspects of the scheme
of the MC modelling of the QCD NLO evolution
not reported in this contribution.
We would like also to point out to some similarities of this
project to works in Refs.~\cite{Kato:1991fs,Tanaka:2003gk}.

\section{NLO PSMC for $C_F^2$ part of NLO nonsinglet kernel}
We start from the ISR ladder of deep inelastic scattering
of the lepton-hadron and our aim is to add 1st order corrections
to the LO vertex in the middle of the ladder, see the Figure below

\begin{figure}[htb]
 {\includegraphics[width=80mm]{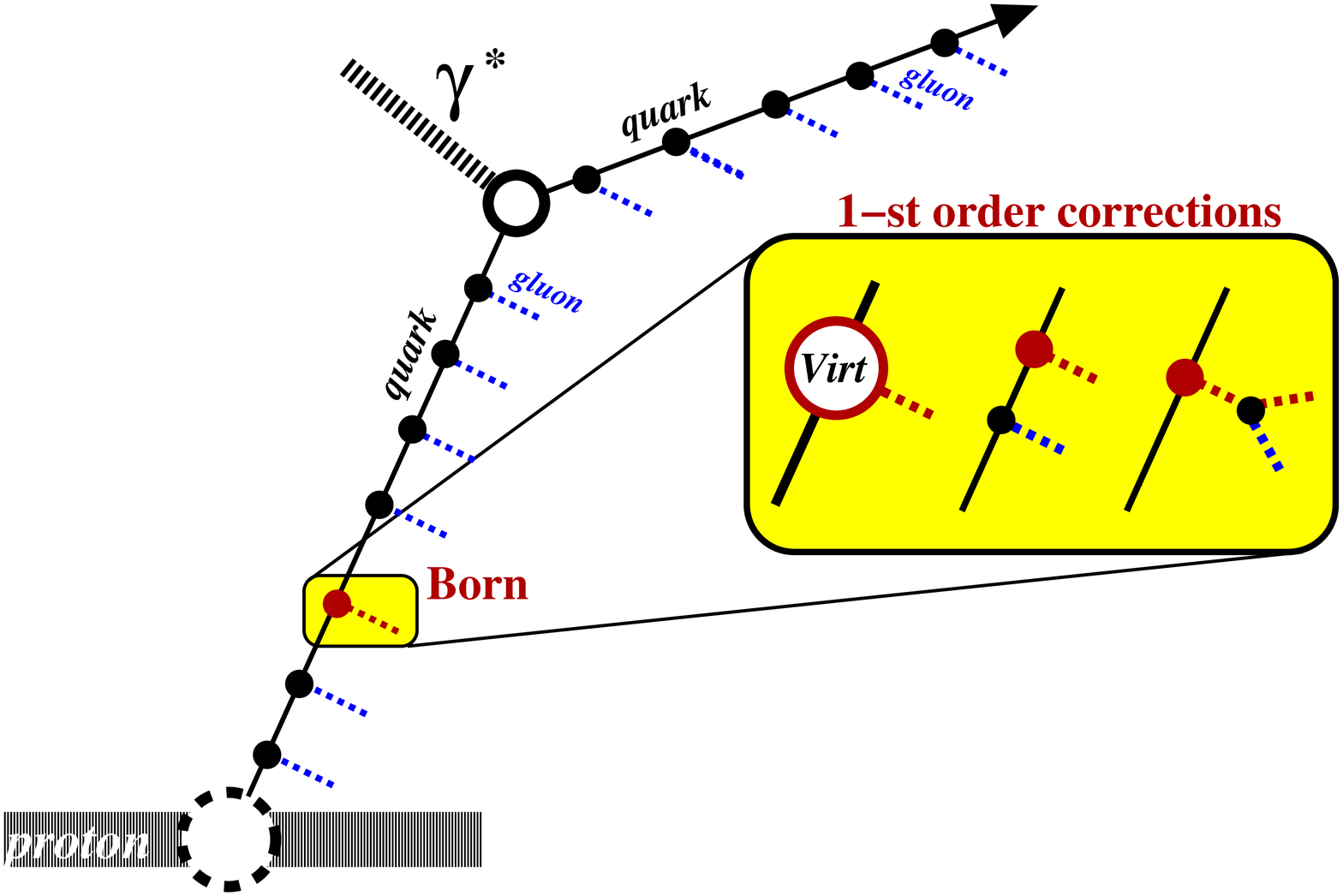}}
\caption{Deep inelastic lepton-proton scattering.}
\label{fig:one}
\end{figure}

\begin{figure}[htb]
\begin{displaymath}
\label{eq:jeden}
 \raisebox{-60pt}{\includegraphics[height=40mm]{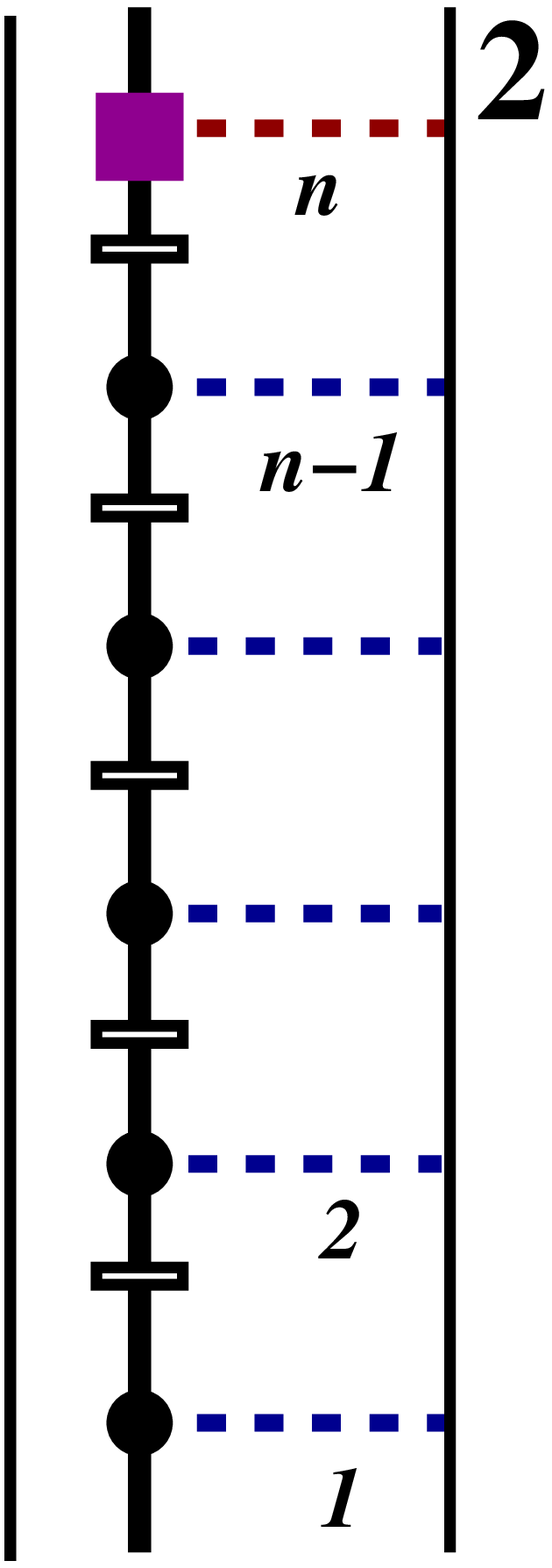}}
+\raisebox{-60pt}{\includegraphics[height=40mm]{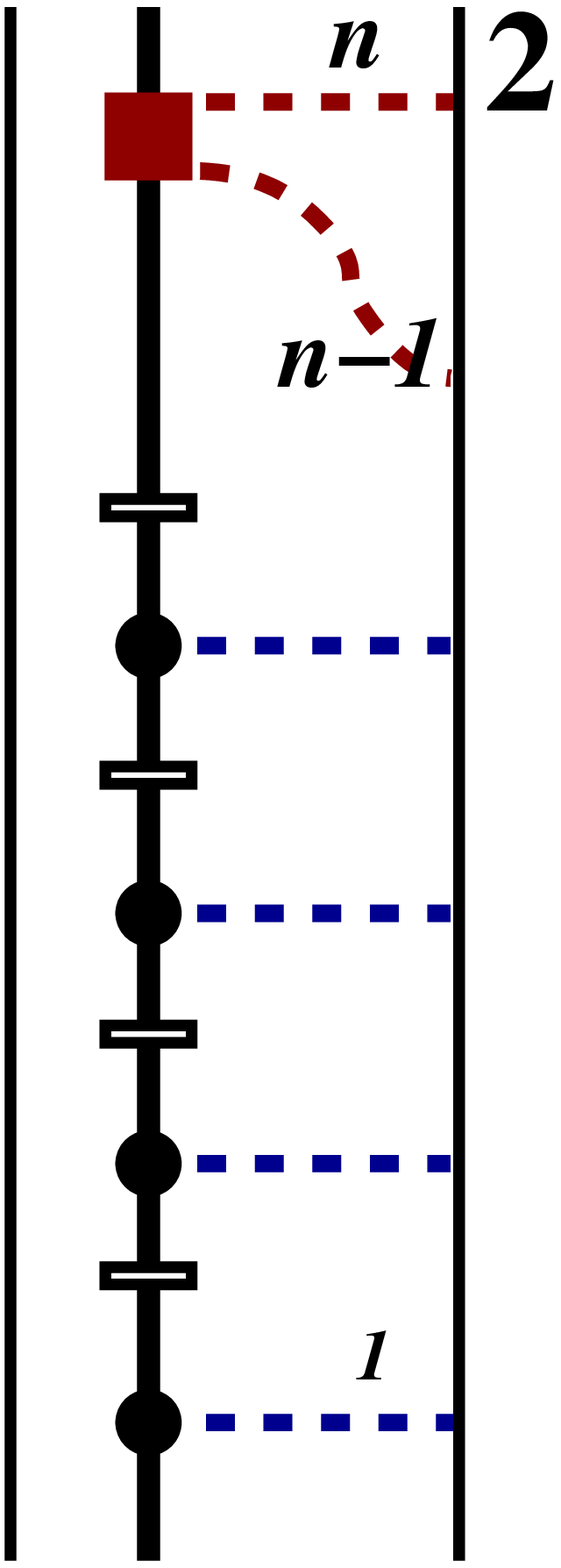}}
+\raisebox{-60pt}{\includegraphics[height=40mm]{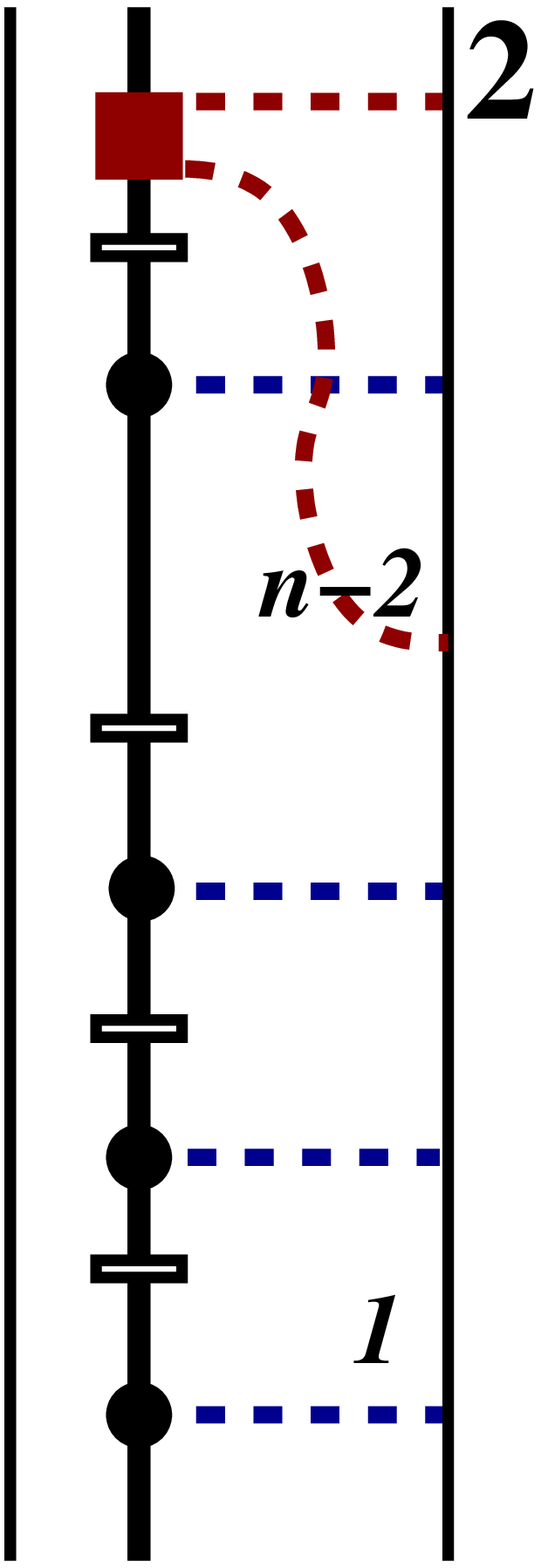}}
+\dots
+\raisebox{-60pt}{\includegraphics[height=40mm]{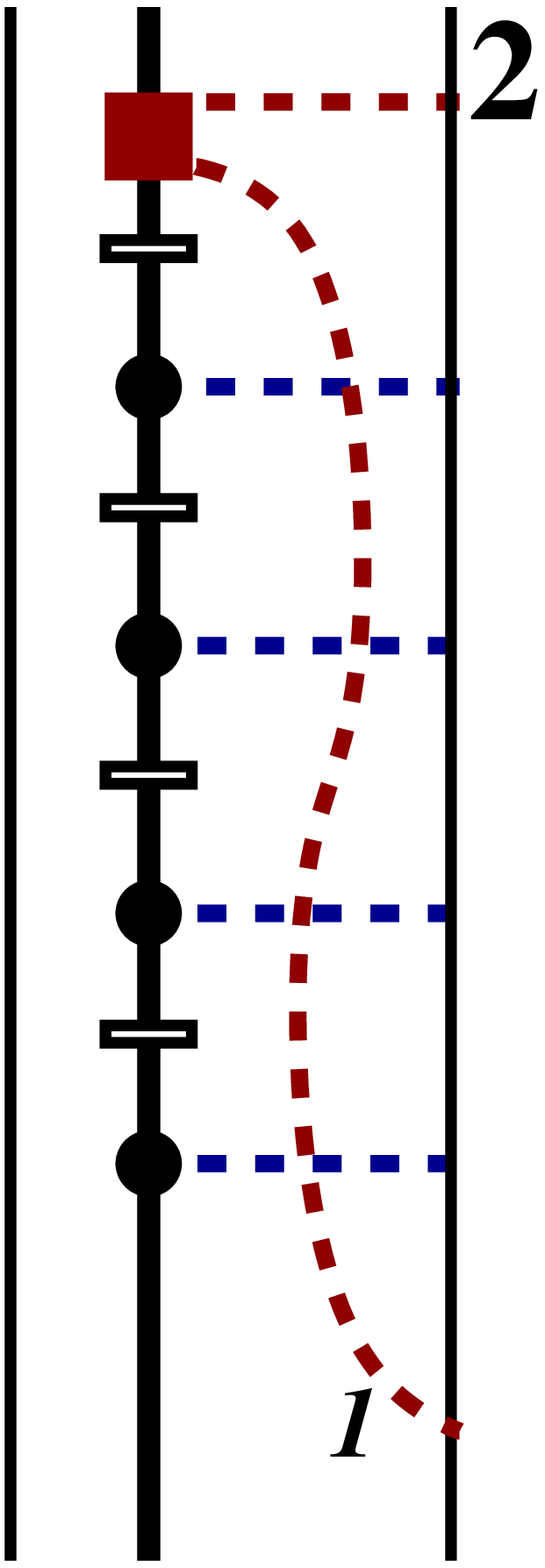}}
\end{displaymath}
\caption{Correcting the last vertex in the LO ladder to NLO level.}
\label{fig:two}
\end{figure}

Let us start with correcting up to NLO level
just one vertex (kernel)
at the end of the ladder (next to hard process)
in the LO PSMC, see the graph in Fig.~\ref{fig:one}.
The multigluon distribution of the LO PSMC
representing the inside of the LO PDF is
\[
e^{-S_{_{ISR}}}
\sum\limits_{n=0}^\infty
\prod\limits_{i=1}^n
\frac{ d^3 k_i}{k_i^0}\;
\theta_{Q>a_i>a_{i-1}}
\rho^{(0)}_{1}(k_i)\;
\delta_{x=\prod z_i},
\]
where $S_{ISR}$ is the Sudakov double log formfactor,
the lightcone variable of the emitted gluon is $\alpha_i=\frac{k^+_i}{2E_h}$,
the angular scale variable is $a_i=\frac{k^T_i}{\alpha_i}$
and the actual distribution of the $q\to Gq$ splitting is
$\rho^{(0)}_{1}(k_i)\;
= \frac{2C_F^2\alpha_s}{\pi} \frac{1}{k_i^{T2}} \frac{1+z_i^2}{2}$,
where
$1-z_i =\alpha_i/(1-\sum_{j=1}^{i-1}\alpha_j)$,
and $C_F$ is colour factor.

In Fig.~\ref{fig:two},
we illustrate again the situation in which the last (top)
vertex is upgraded to NLO level.
Hence in the leftmost graph this vertex includes already 
multiplicative virtual+soft NLO correction:
\[~~~~~~~
\left|
 \raisebox{-10pt}{\includegraphics[height=10mm]{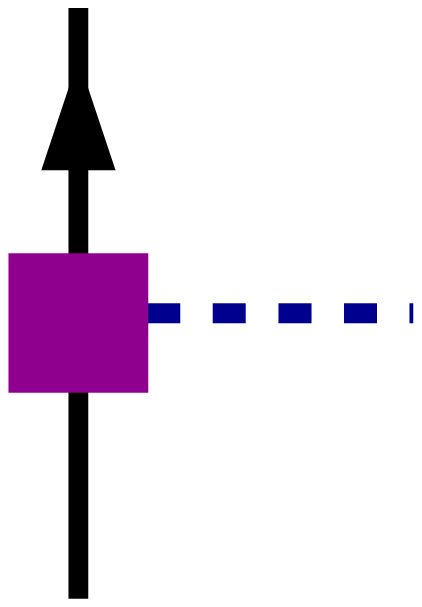}}
\right|^2\!\!\!
=\big(1+2\Re(\Delta_{_{ISR}}^{(1)})\big)\!
\left|
  \raisebox{-10pt}{\includegraphics[height=10mm]{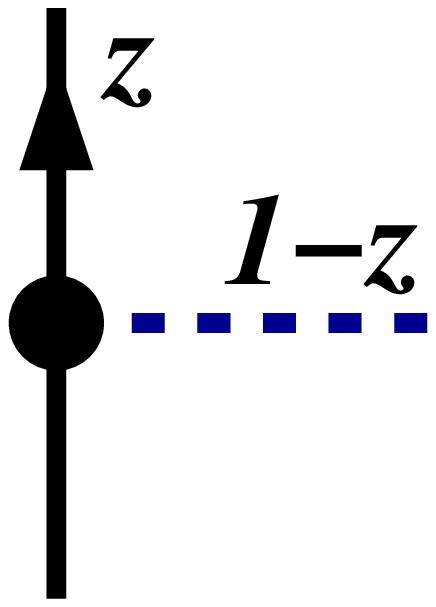}}
\right|^2.
\]
The two-gluon NLO correction coming from two graphs
\raisebox{-8pt}{\includegraphics[height=8mm]{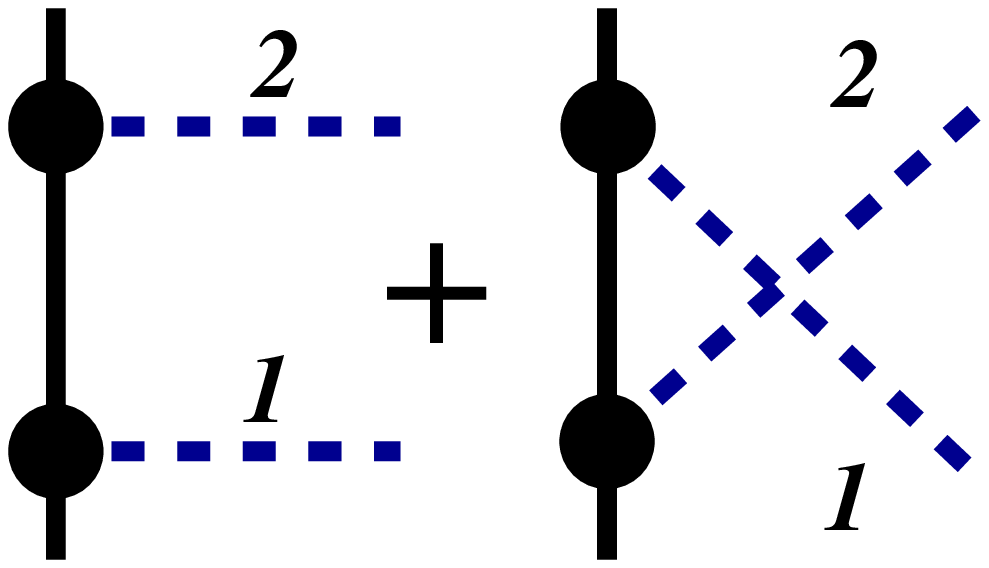}}
is defined as:
\[~~~~~~~
\left|
  \raisebox{-10pt}{\includegraphics[height=10mm]{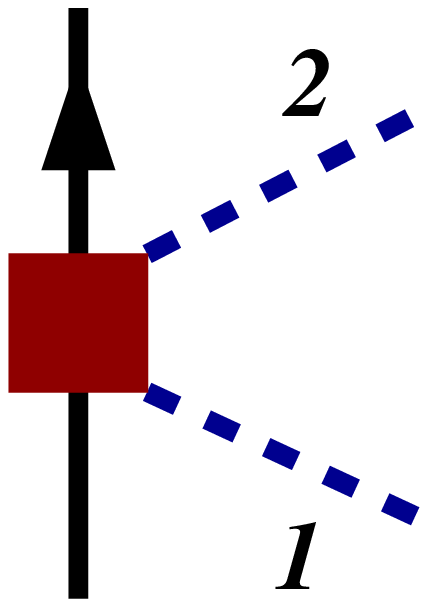}}
\right|^2
=\left|
  \raisebox{-10pt}{\includegraphics[height=10mm]{xBrem2Real.eps}}
\right|^2
-\left|
  \raisebox{-10pt}{\includegraphics[height=10mm]{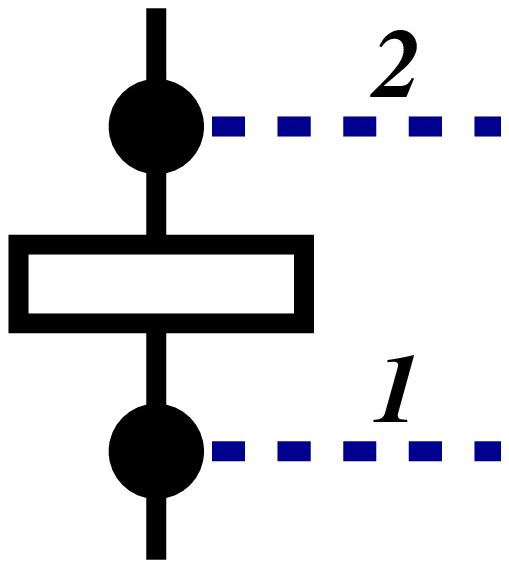}}
\right|^2.
\]
Here, the projection operator ${\bf P}$,
represented by the rectangular box in the graph
\raisebox{-8pt}{\includegraphics[height=8mm]{xBr2ReCt.eps}},
should be understood as operating on the amplitude squared,
spin-summed:
\[~~~~~~~
\left|
  \raisebox{-10pt}{\includegraphics[height=10mm]{xBr2ReCt.eps}}
\right|^2
= \raisebox{-10pt}{\includegraphics[height=10mm]{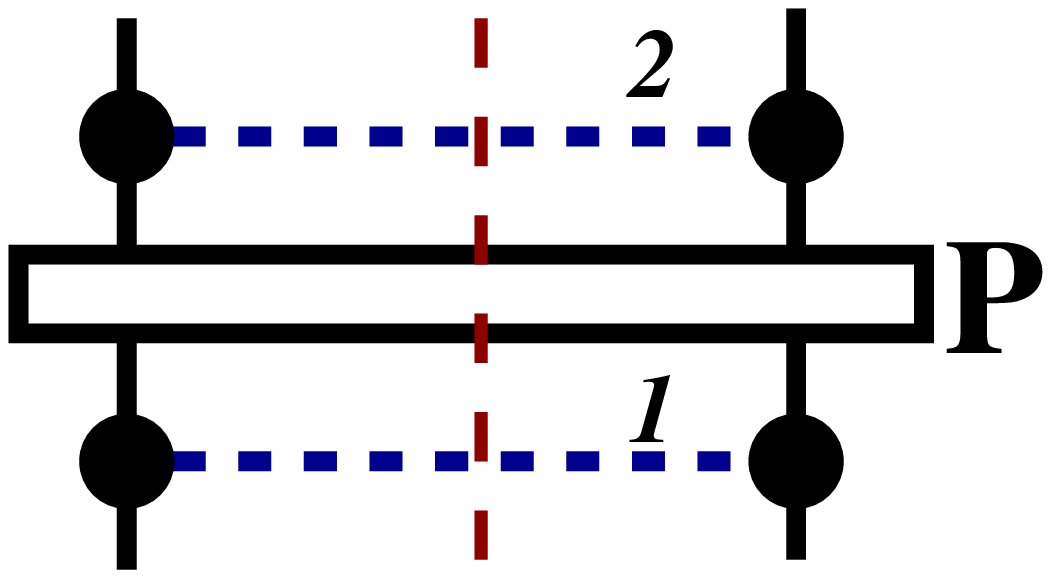}}.
\]
This ${\bf P}$ operator is present in Fig.~\ref{fig:two}
in each ladder many times
and its role is to simplify the exact matrix element to the LO approximation
level all over the phase space.
The role of the NLO correction 
  \raisebox{-8pt}{\includegraphics[height=8mm]{xBrBetISR.eps}}
is to {\em undo} the simplification done by ${\bf P}$, just for 2 gluons.
One could think that it is enough to include two leftmost graphs
of Fig.~\ref{fig:two}.
It is, however, necessary
to include the sum over all {\em LO spectators} down to the beginning
of the ladder, see Fig.~\ref{fig:two},
for the second gluon entering the NLO correction
  \raisebox{-8pt}{\includegraphics[height=8mm]{xBrBetISR.eps}}.

Altogether, the above NLO-corrected multigluon distribution reads as follows
\[
\rho_n(k_i^\mu)=
e^{-S_{_{ISR}}}
\Bigg\{
\raisebox{-35pt}{\includegraphics[height=30mm]{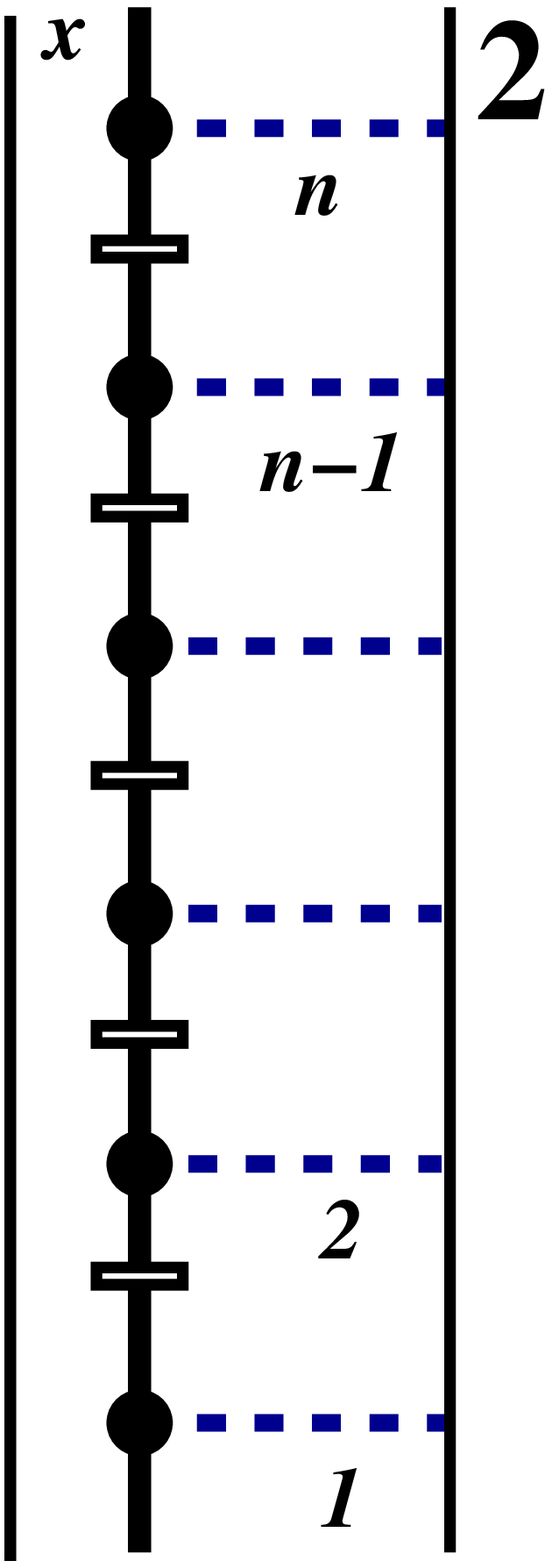}}
+
\raisebox{-35pt}{\includegraphics[height=30mm]{xBrLOa.eps}}
+
\sum\limits_{j=1}^{n-1}
\raisebox{-35pt}{\includegraphics[height=30mm]{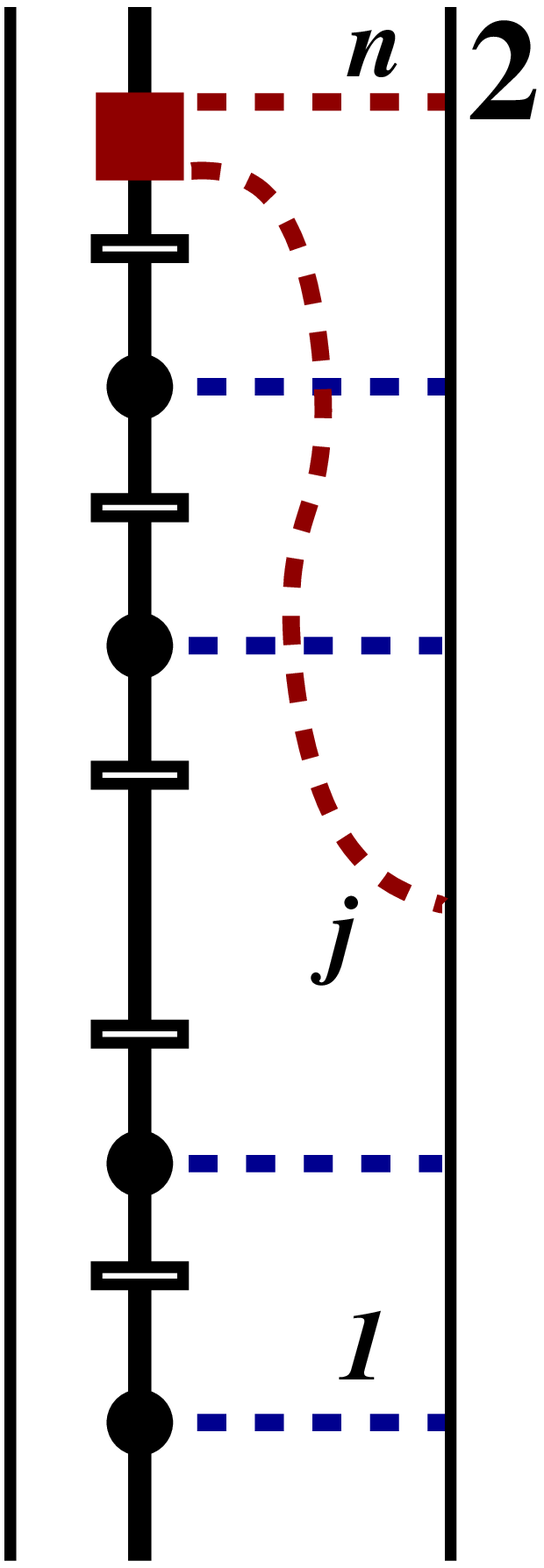}}
\Bigg\}
\]
\[
=e^{-S_{_{ISR}}}
\bigg[ \beta_0^{(1)}(z_n)
+ \sum_{j=1}^{n-1}W(\tilde{k}_n, \tilde{k}_j)
\bigg]
\]
\[~~~\times
    \prod_{i=1}^n\; 
    \theta_{a_i>a_{i-1}}
    \rho^{(1)}_{1}(k_i),
\]
where the MC weight component correcting the 2-gluon distribution is:
\[~~~
W(k_2,k_1)=
\frac{%
 \left| \raisebox{-8pt}{\includegraphics[height=8mm]{xBrBetISR.eps}}
 \right|^2
}{%
 \left| \raisebox{-8pt}{\includegraphics[height=8mm]{xBr2ReCt.eps}}
 \right|^2
}=
\frac{%
 \left| \raisebox{-8pt}{\includegraphics[height=8mm]{xBrem2Real.eps}}
\right|^2
}{%
 \left| \raisebox{-8pt}{\includegraphics[height=8mm]{xBr2ReCt.eps}}
 \right|^2
}\; -1
\]
and the virtual correction is
$
\beta_0^{(1)}=
\frac{%
 \left| \raisebox{-8pt}{\includegraphics[height=8mm]{xBrBet0ISR.eps}}
 \right|^2}%
{ \left|\raisebox{-8pt}{\includegraphics[height=8mm]{xBrBorn.eps}}
 \right|^2
}.
$\\
Amazingly, in the expression for the PDF,
$D(x)=\sum_{n=0}^\infty
\int dLips_n \rho_n(k_1,k_2,...k_n)
\delta(x-\prod_{j=1}^{n} z_j),
$
the integration of the
NLO part $\sum_j W(\tilde{k}_n,\tilde{k}_j)$
can be done analytically leading to:
\[
\sum_{n=1}^\infty\;
\int du
\int\limits_{Q>a_n>a_{n-1}}\!\!\!\!\!\!
\frac{da_n}{a_n}\;
\Peu^{(1)}_{qq}(u)\;
\]
\[~~~~~\times
\bigg( \prod_{i=1}^{n-1}\; 
    \int\limits_{a_{i+1}>a_i>a_{i-1}}\!\!\!\!\!\!
\frac{da_i}{a_i} \Peu^{(0)}_{qq}(z_i)
\bigg)
\delta_{x=u\prod_{j=1}^{n-1} z_j},
\]
recovering for the last emission precisely the
NLO part (including virtuals) of standard DGLAP kernel
$\Peu^{(1)}_{qq}(u)$
defined according to:
\[
\Peu^{(1)}_{qq}(u)
\ln\frac{Q}{q_0}=\!\!\!\!\!\!\!
\int\limits_{Q>a_n>a_{0}}\!\!\!\!\!\!  d^3\eta_n\;
    \rho^{(1)}_{1B}(k_n)\;
    \beta_0^{(1)}(z_n)
  \delta_{u=z_n}
\]
\[~~
+\!\!\!\!\!
\int\limits_{Q>a_{n}>a_{0}}\!\!\!\!\!\!\!\!\!\!  d^3\eta_{n}\;
\int\limits_{a_{n}>a_{n'}>0}\!\!\!\!\!\!\!\! d^3\eta_{n'}\;
 \beta_1^{(1)}(\tilde{k}_{n}, \tilde{k}_{n'})\;
  \delta_{u=z_n z_{n'}}.
\]
In this way at the inclusive level
the NLO standard inclusive kernel of DGLAP is truly reproduced
for the last ($n$-th) vertex in the ladder.

One can repeat the same procedure for any
vertex number $p_1$ in the middle
of the ladder, sum up over $p_1$, then apply the same for
any two vertices $p_1$ and $p_2$ and so on, such that finally
all vertices in the ladder are at the NLO level.
The above general procedure will yield the following multigluon
distribution:
\[
\rho_n(k_l)=
e^{-S_{_{ISR}}}
\Bigg\{
\raisebox{-30pt}{\includegraphics[height=30mm]{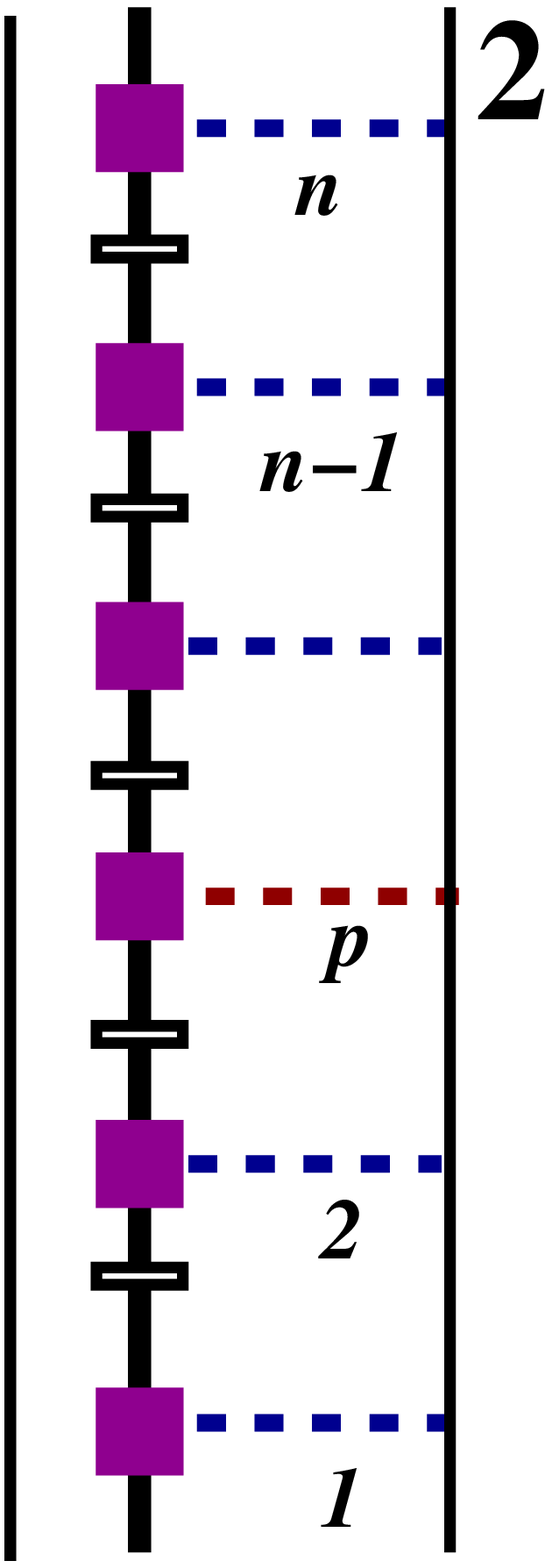}}
+
\sum\limits_{p_1=1}^{n}
\sum\limits_{j_1=1}^{p_1-1}
\raisebox{-30pt}{\includegraphics[height=30mm]{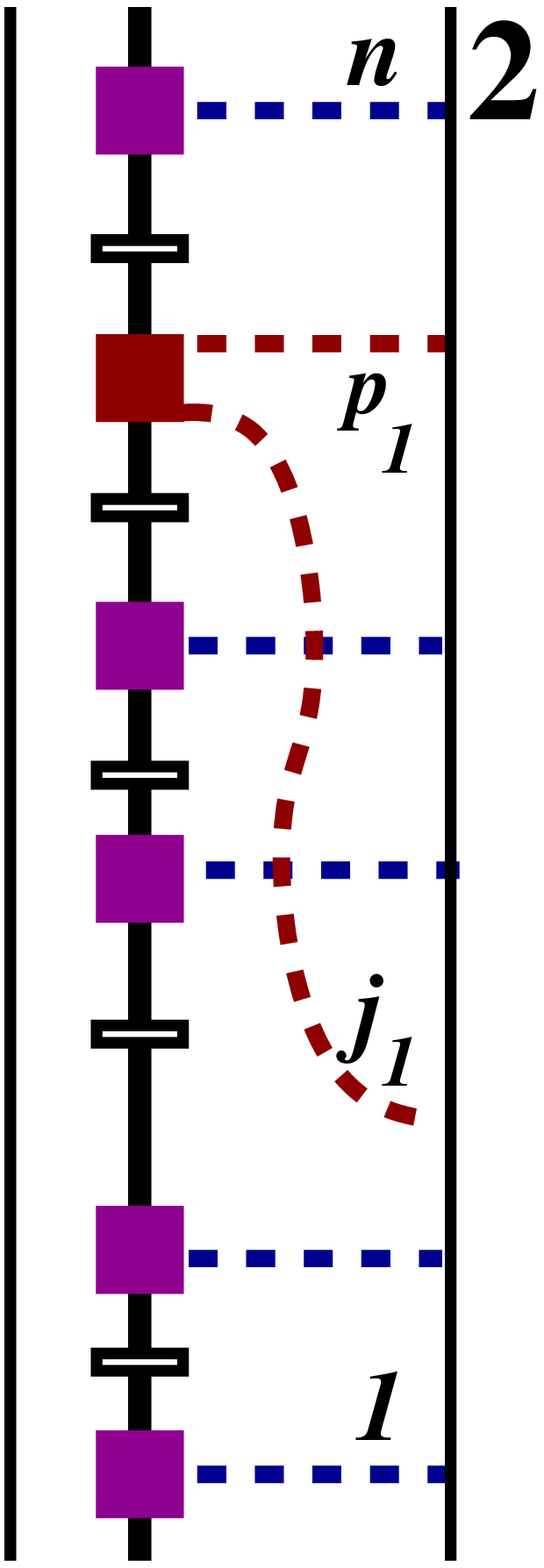}}
\]
\[~~
+
\sum\limits_{p_1=1}^{n}
\sum\limits_{p_2=1}^{p_1-1}
\sum\limits_{j_1=1 \atop j_1\neq p_2}^{p_1-1}
\sum\limits_{j_2=1 \atop j_2\neq p_1,j_2}^{p_2-1}
\raisebox{-30pt}{\includegraphics[height=30mm]{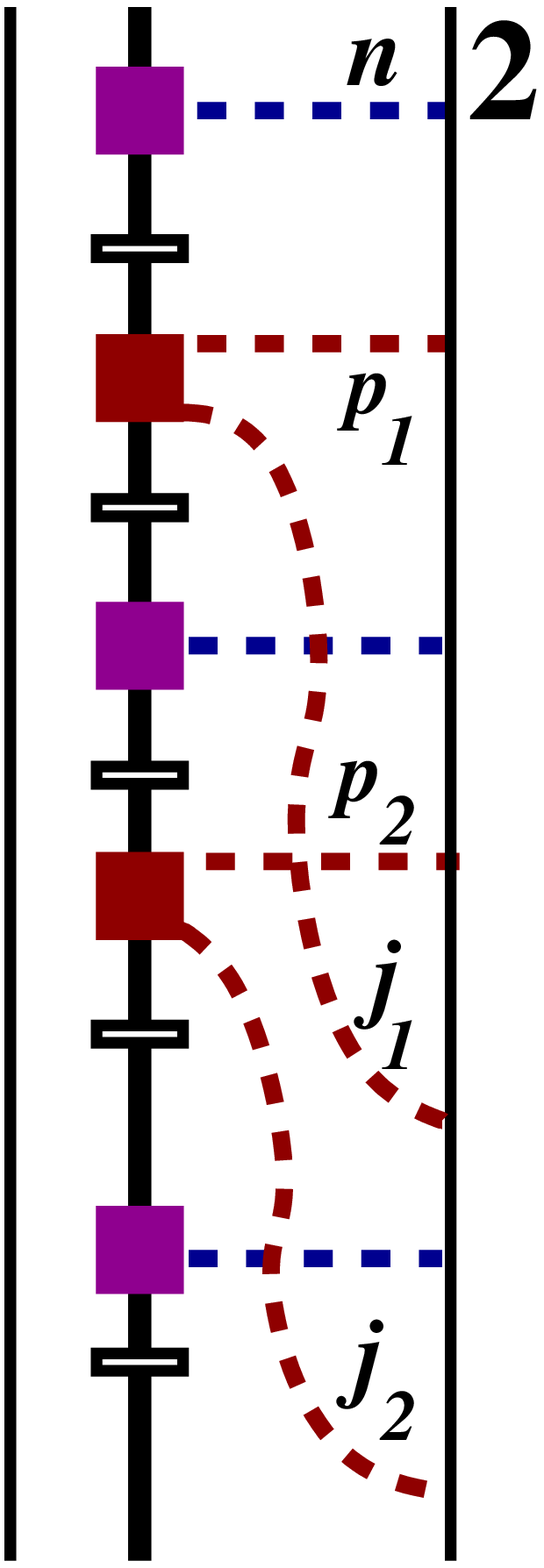}}
\Bigg\}
=
\]
\[
=e^{-S_{_{ISR}}}
\bigg[
\beta_0^{(1)}(z_p)
+ \sum\limits_{p=1}^{n} \sum_{j=1}^{p-1}W(\tilde{k}_p, \tilde{k}_j)
\]
\[~~~
+\sum\limits_{p_1=1}^{n}
\sum\limits_{p_2=1}^{p_1-1}
\sum\limits_{j_1=1 \atop j_1\neq p_2}^{p_1-1}
\sum\limits_{j_2=1 \atop j_2\neq p_1,j_2}^{p_2-1}
\!\!\!
W(\tilde{k}_{p_1}, \tilde{k}_{j_1})
W(\tilde{k}_{p_2}, \tilde{k}_{j_2})
\]
\begin{equation}
\label{eq:dwa}
~~~
+\dots
\bigg]
\prod_{i=1}^n\; \theta_{a_i>a_{i-1}}
    \rho^{(1)}_{1}(k_i)
    \beta_0^{(1)}(z_i).
\end{equation}
The above formula was already tested numerically
with three digit precision.
\begin{center}
  {\includegraphics[width=80mm,height=50mm]{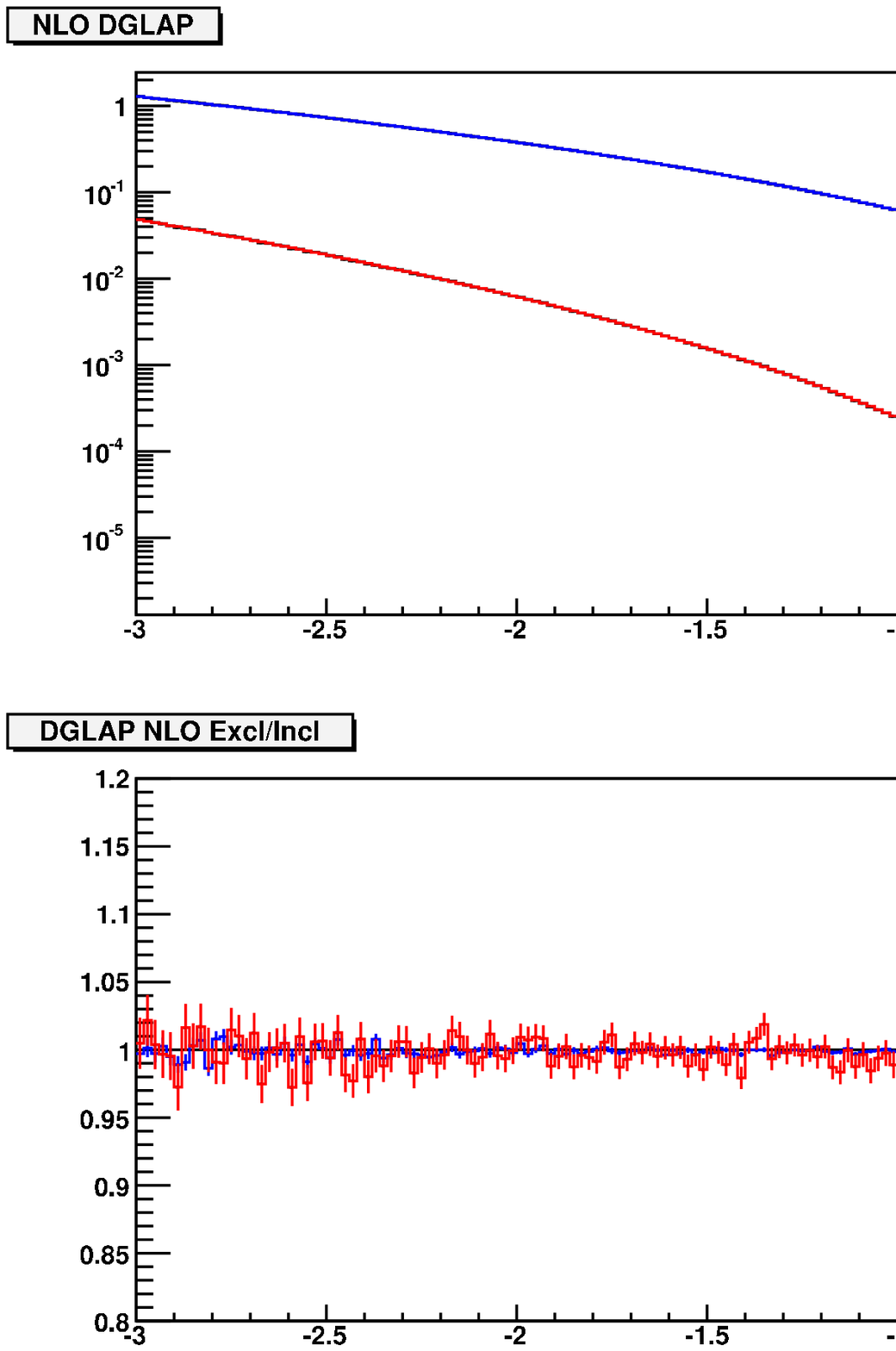}}
\end{center}
In the upper part of the figure above we present
numerical results for the (inclusive) PDF, $D(x,Q)$, from
two Monte Carlo models, 
one with traditional inclusive evolution kernels
and another one implementing
our new NLO distribution of Eq.~(\ref{eq:dwa}).
The trivial LO contribution is excluded from the comparison
and the two curves represent the term in Eq.~(\ref{eq:dwa})
with single $W$ and with double $W$.
Since the results cannot be distinguished in the upper plot,
we also show their ratio in the lower plot in the Figure above.
The evolution runs from $Q_0=10$GeV to $Q=1$TeV, starting from 
$D(x,Q_0)=\delta(1-x)$.
The ratio demonstrates 3-digit agreement, in units of the LO.

Let us note that the Monte Carlo weight
implementing NLO corrections is positive,
with small dispersion and without any nasty tails.

\section{Adding $C_FC_A$ part of the NLO nonsinglet kernel of NLO PSMC }
Straightforward inclusion of the gluon pair diagram
in the method of the previous section
would be mathematically correct,
but the Monte Carlo weight would be spoilt
due to presence of Sudakov double logarithmic
contribution $+S_{FSR}$ in the 2-real-gluon correction
\begin{displaymath}
\left|
  \raisebox{-9pt}{\includegraphics[height=9mm]{xBrBetISR.eps}}
\right|^2
=\left|
  \raisebox{-9pt}{\includegraphics[height=9mm]{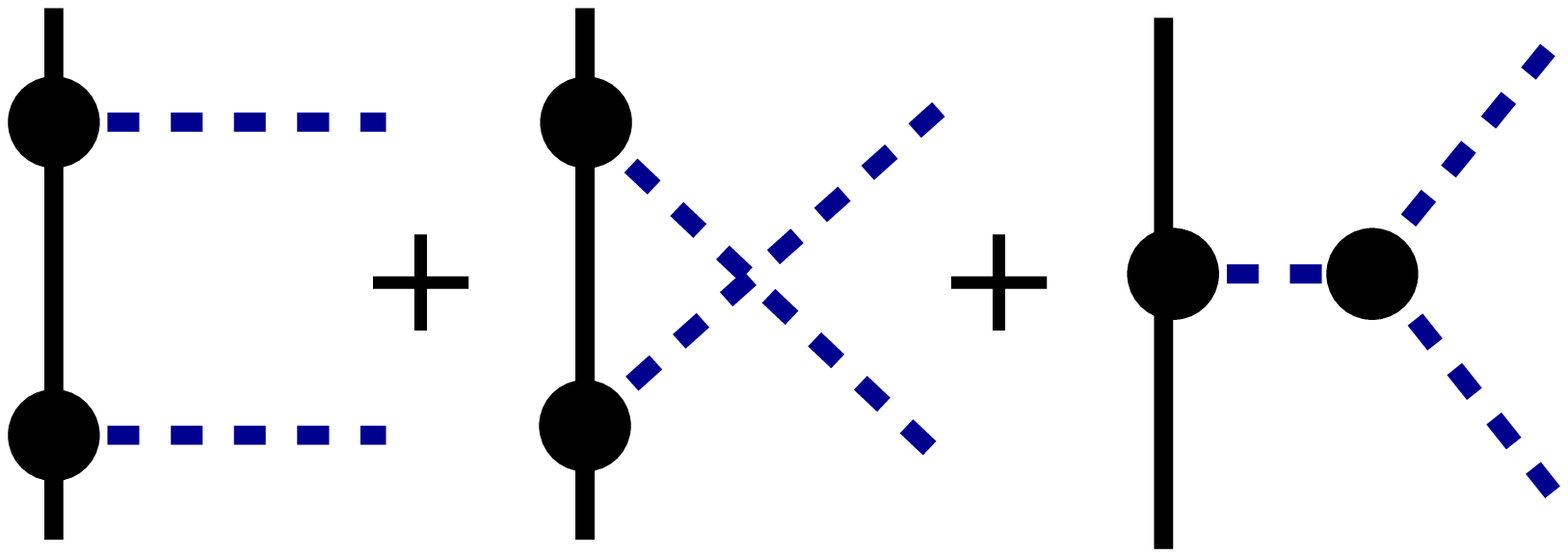}}
\right|^2
-\left|
  \raisebox{-9pt}{\includegraphics[height=9mm]{xBr2ReCt.eps}}
\right|^2
\end{displaymath}
compensated by $-S_{FSR}$ component in the virtual correction
\begin{displaymath}
\left|
 \raisebox{-9pt}{\includegraphics[height=9mm]{xBrBet0ISR.eps}}
\right|^2
=\big(1+2\Re(\Delta_{_{ISR}}+V_{_{FSR}})\big)
\left|
  \raisebox{-9pt}{\includegraphics[height=9mm]{xBrBorn.eps}}
\right|^2.
\end{displaymath}
Resummation/exponentiation of FSR is therefore mandatory.
Let us start again with the problem of upgrading to the NLO level
the last vertex of the LO ladder.
In the MC weight we add an extra term, which in
the graphical form looks as follows
\begin{displaymath}
e^{-S_{_{ISR}}-S_{_{FSR}}}
 \sum\limits_{n,m=0}^{\infty}
 \sum\limits_{r=1}^{m}
\raisebox{-20pt}{\includegraphics[height=20mm]{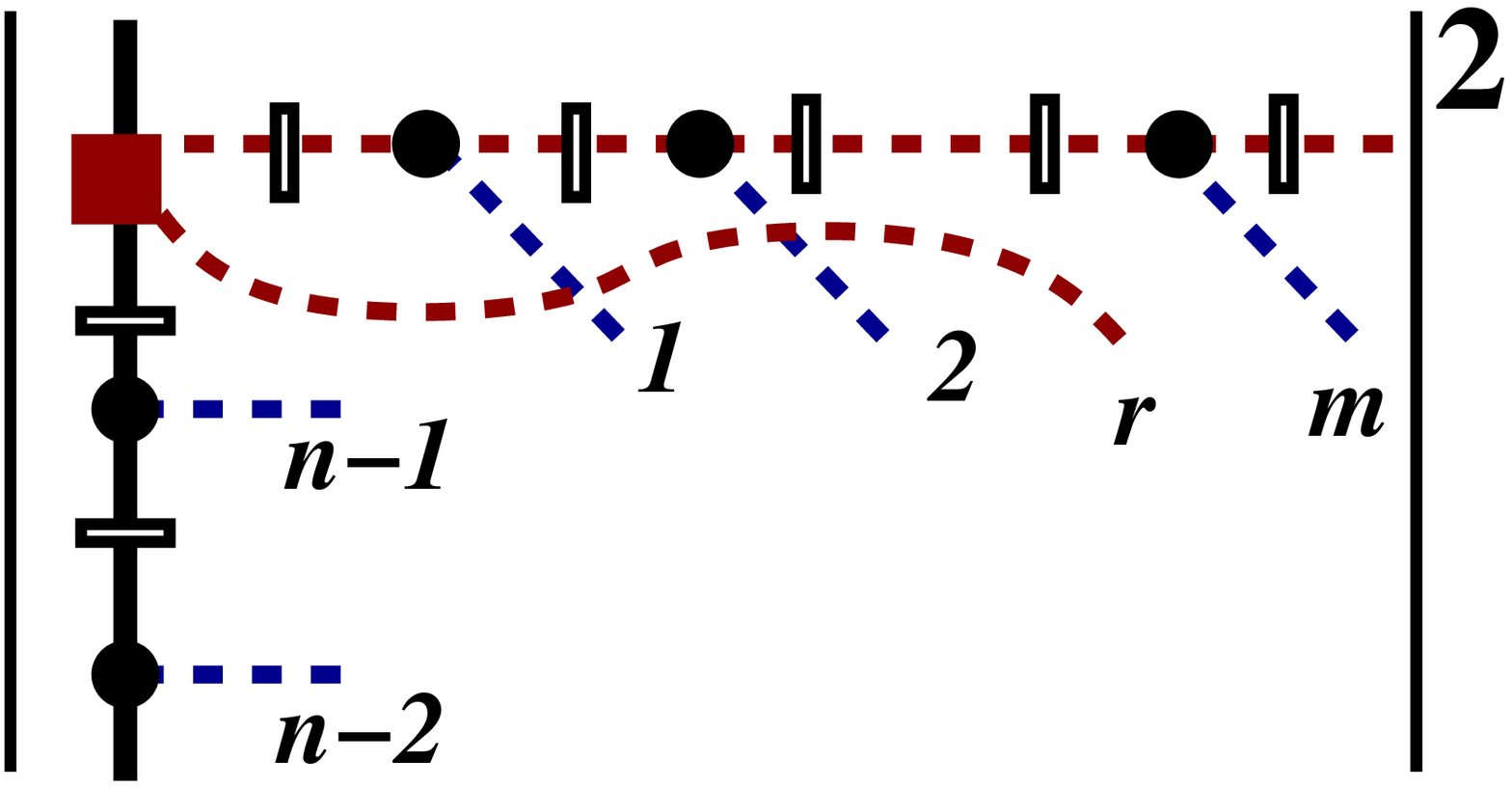}},
\end{displaymath}
where the Sudakov term $S_{_{FSR}}$ is subtracted in the virtual part:
\begin{displaymath}
\left|
 \raisebox{-9pt}{\includegraphics[height=9mm]{xBrBet0ISR.eps}}
\right|^2
=\big(1+2\Re(\Delta_{_{ISR}}+V_{_{FSR}} {-S_{_{FSR}}})\big)
\left|
  \raisebox{-9pt}{\includegraphics[height=9mm]{xBrBorn.eps}}
\right|^2,
\end{displaymath}
and the FSR soft counterterm is subtracted in the 2-real-gluon part:
\begin{displaymath}
\left|
  \raisebox{-8pt}{\includegraphics[height=8mm]{xBrBetISR.eps}}
\right|^2
\!\!\!
=\left|
  \raisebox{-8pt}{\includegraphics[height=8mm]{xBrem2gReal.eps}}
\right|^2\!\!\!
-\left|
  \raisebox{-8pt}{\includegraphics[height=8mm]{xBr2ReCt.eps}}
\right|^2
-\left|
  \raisebox{-8pt}{\includegraphics[height=8mm]{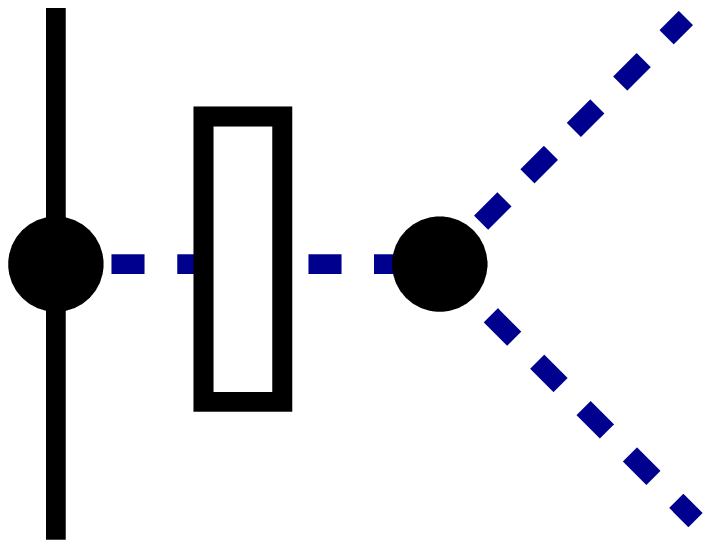}}
\right|^2\!\!.
\end{displaymath}
It is remarkable that
both above NLO corrections are free of any soft and collinear
divergences, in spite of their complicated structure.
This is thanks to a clever choice of the FSR counterterm
\raisebox{-8pt}{\includegraphics[height=8mm]{xBr2gReCt.eps}},
which is iterated in the FSR LO MC for the gluon emitted
from the ladder.
It was helpful in the above exercise that we were
employing angular ordering in the LO MC both for ISR and FSR parts.

Altogether, the complete MC distribution, with
the last LO vertex in the ladder upgraded to
the NLO level looks as follows:
\[
\rho_{n,m}^{[1]}(k^\mu_l,k'^\mu_{l'})=
e^{-S}
\Bigg\{
\raisebox{-25pt}{\includegraphics[height=25mm]{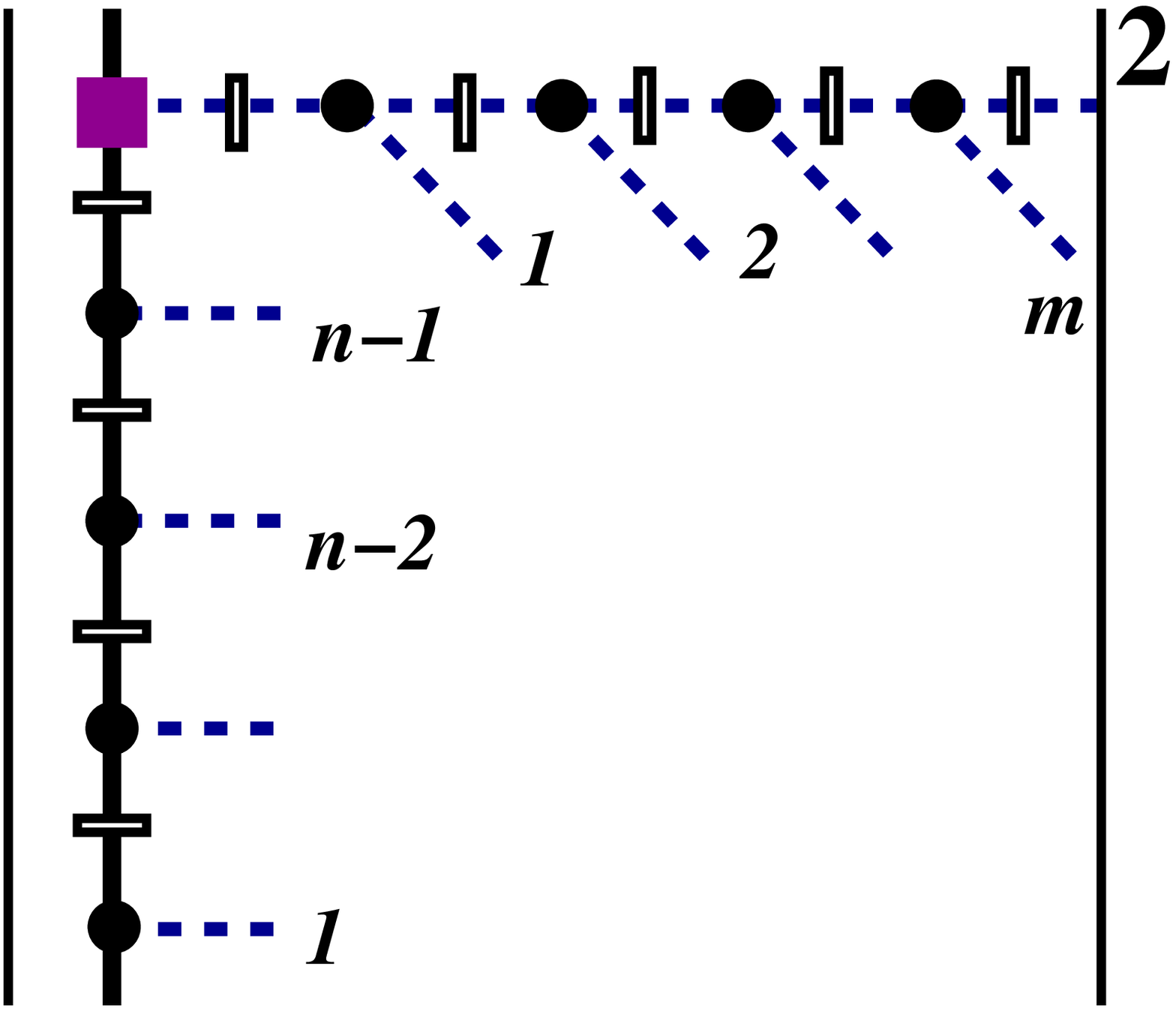}}+
\]
\[
+\sum\limits_{j=1}^{n-1}
\raisebox{-25pt}{\includegraphics[height=25mm]{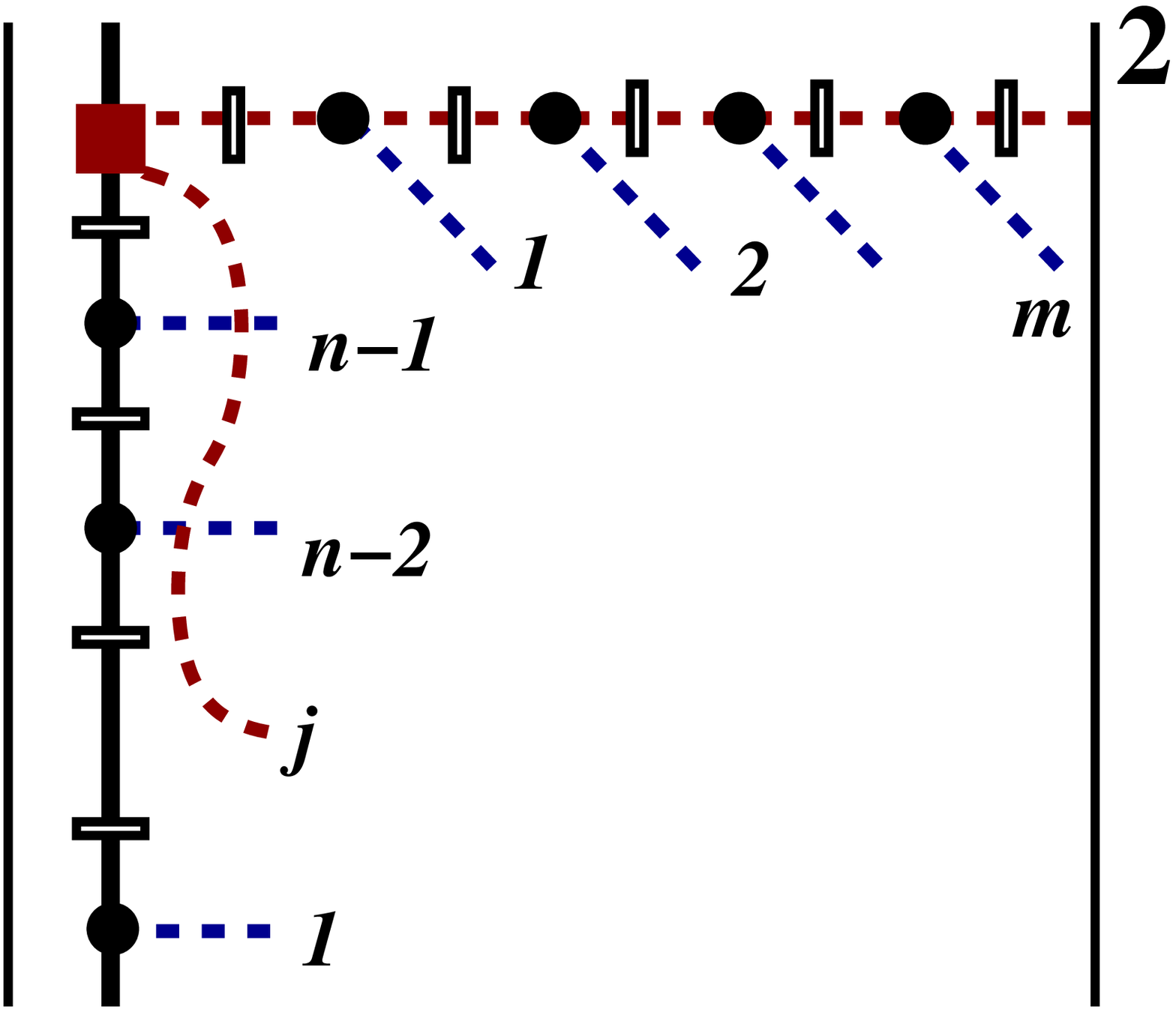}}
\!\!\!\!
+\sum\limits_{r=1}^{m}
\raisebox{-25pt}{\includegraphics[height=25mm]{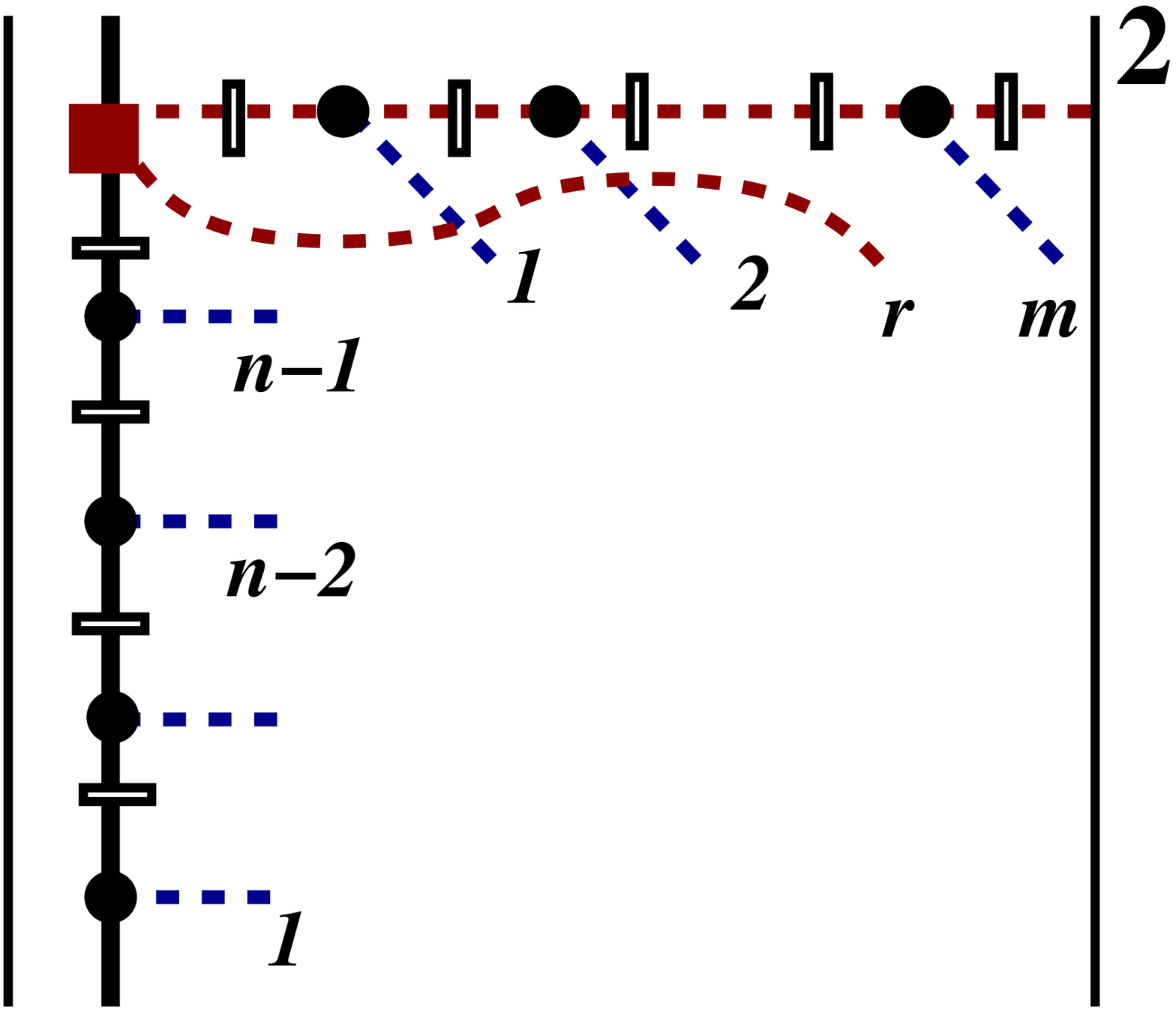}}
\!\!
\Bigg\}
\]
\[
=
\bigg( \prod_{i=1}^n\; 
    \rho^{(1)}_{1}(k_i)
    \theta_{a_i>a_{i-1}}
\bigg)
\]
\[\times
\bigg( \prod_{j=1}^m\; 
    \rho^{(1)}_{1V}(k'_j)
    \theta_{a_{nj}>a_{n(l-1)}}
\bigg)
e^{-S_{_{ISR}}-S_{_{FSR}}}
\]
\[\times
\bigg[
\beta_0^{(1)}(z_n)
+\sum_{j=1}^{n-1} W(\tilde{k}_n, \tilde{k}_j)
+\sum_{r=1}^{m}   W(\tilde{k}_n, \tilde{k}'_r)
\bigg],
\]
where NLO correction building blocks are
\[
\beta_0^{(1)}\equiv
\frac{%
 \left| \raisebox{-7pt}{\includegraphics[height=7mm]{xBrBet0ISR.eps}}
 \right|^2}%
{ \left|\raisebox{-7pt}{\includegraphics[height=7mm]{xBrBorn.eps}}
 \right|^2
},\quad
W(k_2,k_1)\equiv
\frac{%
 \left| \raisebox{-7pt}{\includegraphics[height=7mm]{xBrBetISR.eps}}
 \right|^2
}{%
 \left| \raisebox{-7pt}{\includegraphics[height=7mm]{xBr2ReCt.eps}}
 \right|^2
+\left| \raisebox{-7pt}{\includegraphics[height=7mm]{xBr2gReCt.eps}}
\right|^2
}
\]
\[~~~~~~~~~~~~~~~
=
\frac{%
 \left| \raisebox{-7pt}{\includegraphics[height=7mm]{xBrem2gReal.eps}}
\right|^2
}{%
 \left| \raisebox{-7pt}{\includegraphics[height=7mm]{xBr2ReCt.eps}}
 \right|^2
+\left| \raisebox{-7pt}{\includegraphics[height=7mm]{xBr2gReCt.eps}}
\right|^2
}\; -1.
\]
Again it is possible to check analytically, that the above apparently
complicated MC weight reproduces {\em exactly} 
the traditional NLO integrated kernels for the last vertex in the ladder.
We have also performed numerical tests in which the above
distribution is modelled for the $n=1,2$ ISR gluons and any number
of FSR gluons.
The comparison between MC and analytical result is in the following
plot:
\begin{center}
{\includegraphics[width=80mm,height=50mm]{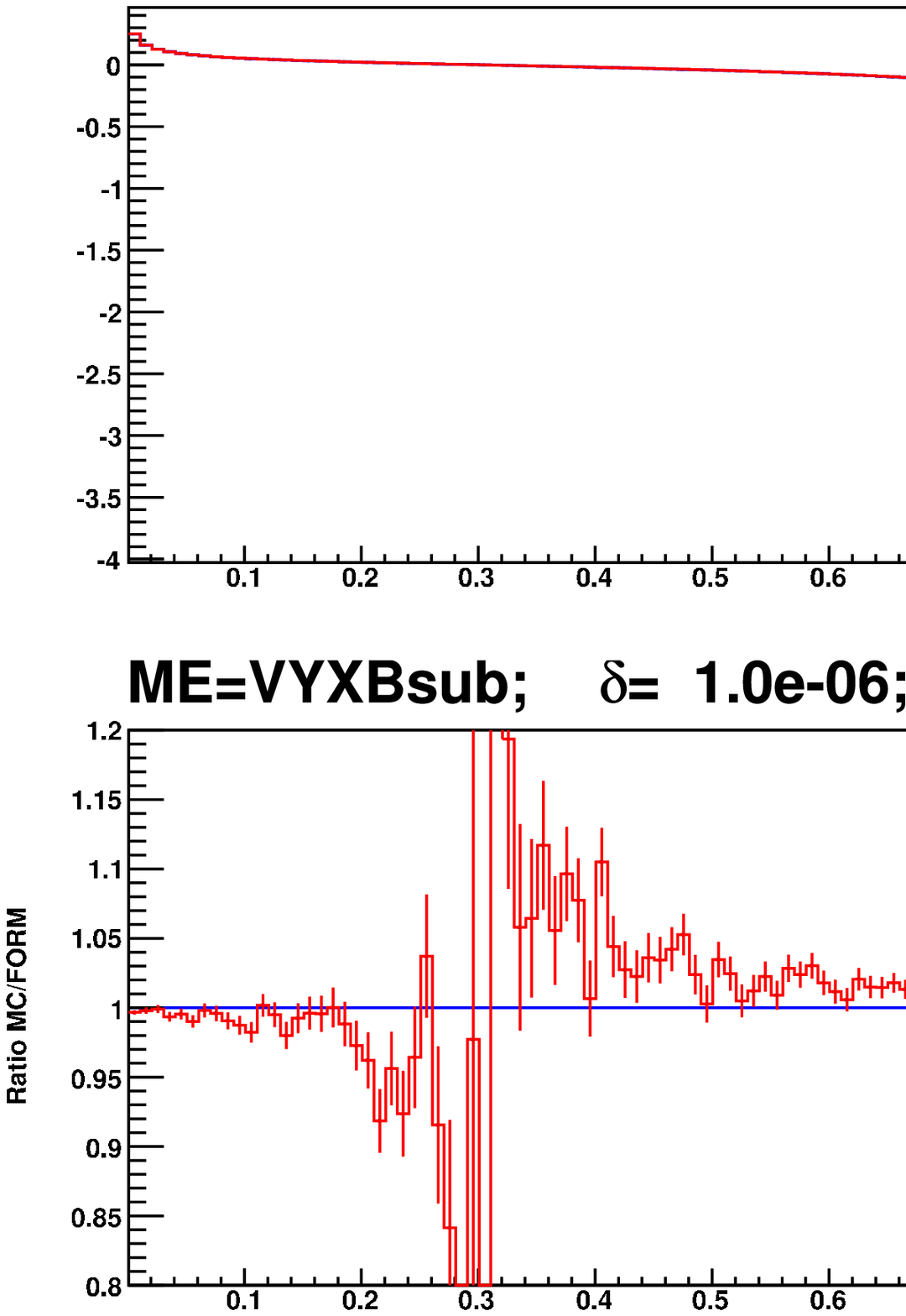}}
\end{center}
It shows again the excellent agreement
of the exclusive implementation of the NLO
corrections in the PSMC and the analytical crosscheck, this time
for the complete gluonstrahlung,
including both $C_F^2$ and $C_FC_A$ parts.
Again we have checked that the MC weight is positive and the
distribution of the MC weights is narrow.

\section{Summary and prospects}
Summarizing we state that the
first serious {\em feasibility study} of the true NLO exclusive MC
parton shower is almost complete for the non-singlet NLO DGLAP.
The building block for the corresponding parton shower MC is
tested numerically.
Further work will cover the following areas:
\begin{itemize}
\item
  Short range aim: Complete non-singlet, also for hard process.
\item
  Middle range aim: Complete singlet (quark-gluon transitions).
\item
  Optimize MC weight evaluation (CPU time).
\item
  Complete NLO MC for DIS at HERA and W/Z production at LHC.
\item
  Interface to standard NLO $\overline{MS}$ PDFs,
\item
  Extensions towards CCFM/BFKL.
\end{itemize}

\section{Acknowledgements}
This research has been partly supported by the the Polish Ministry of 
Science and Higher Education grants  No. 1289/B/H03/2009/37
and No.\ 153/6.PR UE/2007/7.


\end{document}